\documentclass[twocolumn,prl]{revtex4}
\usepackage{graphicx}
\usepackage{amsmath}

\def\eq#1{Eq.~(\ref{#1})}
\def\fig#1{Fig.~\ref{#1}}

\begin{document}

\title{Effect of Poisson ratio on cellular structure formation}
\author{I. B. Bischofs}
\author{U. S. Schwarz}
\affiliation{Interdisciplinary Center for Scientific Computing, University of Heidelberg, 
Im Neuenheimer Feld 368, D-69120 Heidelberg, Germany}
\affiliation{Max Planck Institute of Colloids and Interfaces, D-14424 Potsdam, Germany}

\begin{abstract}
  Mechanically active cells in soft media act as force dipoles. The
  resulting elastic interactions are long-ranged and favor the
  formation of strings. We show analytically that due to screening,
  the effective interaction between strings decays exponentially, with
  a decay length determined only by geometry. Both for disordered and
  ordered arrangements of cells, we predict novel phase transitions
  from paraelastic to ferroelastic and anti-ferroelastic phases as a
  function of Poisson ratio.
\end{abstract}

\maketitle

Predicting structure formation and phase behaviour from the
microscopic interaction laws is a formidable task in statistical
mechanics, especially if the interaction laws are long-ranged or
anisotropic.  In biological systems, the situation is further
complicated because interacting components are active in the sense
that informed by internal instructions (e.g.\ genetic programmes for
cells) and fueled by energy reservoirs (e.g.\ ATP), they react to
input signals in a complicated way, which usually does not follow from
an energy functional. Therefore these systems are often described by
stochastic equations \cite{c:vics95,c:schw98}. One drawback of this
approach is that typically the stochastic equations have to be
analyzed by numerical rather than analytical methods.  However, for
specific systems structure formation of active particles can be
predicted from extremum principles. In this case, analytical progress
might become feasible again, in particular if analogies exist to
classical systems of passive particles.  One example of this kind
might be hydrodynamic interactions of active particles like swimming
bacteria \cite{c:hatw04}. Here this is demonstrated for another
example, namely mechanically active cells interacting through their
elastic environment \cite{uss:bisc03a,uss:bisc04a}.

Our starting point is the observation that generation and propagation
of elastic fields for active particles proceed in a similar way as
they do for passive particles like defects in a host crystal, e.g.\
hydrogen in metal \cite{e:wagn74,uss:schw02a}. For a local
force distribution in the absence of external fields, the overall
force (monopole) applied to the elastic medium vanishes due to
Newton's third law \cite{c:mann01}.  Therefore each particle is
characterized in leading order of a multipolar expansion by a force
dipole tensor $P_{ij}$. For many situations of interest, including
cells in soft media, this force dipole will be anisotropic and can be
written as $P_{ij} = P n_i n_j$, where $\vec n$ is the unit vector
describing particle orientation and $P$ is the force dipolar moment.
The perturbation of the surrounding medium resulting from a force
dipole $P'_{kl}$ positioned at $\vec{r'}$ is described by the strain
tensor $u_{ij}(\vec{r}) = \partial_{j} \partial'_{l}
G_{ik}(\vec{r},\vec{r'}) P'_{kl}$, where summation over repeated
indices is implied and $G_{ij}$ is the Green function for the given
geometry, boundary conditions and material properties of the
sample. The strain $u_{ij}$ generated by one particle causes a
reaction of another particle leading to elastic interactions.
The essential difference between active and passive particles is that
cells respond to strain in an opposite way as do defects. For defects
the interaction with the environment leads to a linearized potential
$V= - P_{ij} u_{ij}$ \cite{e:wagn74}. For example, an anisotropic
defect attracting the atoms of its host lattice turns away from
tensile strain, in this way enhancing the displacement fields in the
medium. In contrast, for mechanically active cells like fibroblasts,
experimental observations suggest that they adopt positions and
orientations in such a way as to effectively minimize the scalar
quantity $W=P_{ij}u_{ij}=-V$ \cite{uss:bisc03a,uss:bisc04a}.  For
tensile strain, this implies that contractile cells actively align
with the external field.  Because they pull against it, they reduce
displacement.  For fibroblast-like cells, this behaviour might have
evolved in the context of wound healing, when cell traction is
required to close wounds.  For a translationally invariant system, the
effective interaction potential between elastically interacting cells
therefore follows as
\begin{equation} \label{eq:two_particle_interaction}
W = P_{ij} u_{ij} = - P_{ij} \partial_{j} \partial_{l} G_{ik}(\vec{r}-\vec{r'}) P'_{kl},
\end{equation}
where we have used $\partial'_{l} G_{ik} = - \partial_{l} G_{ik}$.  In
general, $W$ scales as $1/r^3$ with distance $r$ and has a non-trivial
angular dependance.

The simplest model for the elastic properties of the extracellular
environment is isotropic linear elasticity. Thus there are two elastic
constants: the Young modulus $E$ describes the rigidity of the
material and the Poisson ratio $\nu$ the relative importance of
compression and shear.  Its maximal value is $\nu = 1/2$
(incompressible material). If such a material is tensed in one
direction, the shear mode dominates and it contracts in the
perpendicular directions (\textit{Poisson effect}). For common
materials, the minimal value for the Poisson ratio is $\nu = 0$. Then
the compression mode dominates and uniaxial tension does not translate
into lateral contraction. Isotropic linear elasticity is a reasonable
assumption for the synthetic polymer substrates commonly used to study
mechanical effects in cell culture \cite{c:demb99,uss:schw02b}. In the
following, we therefore use the Boussinesq Green function for
particles exerting tangential forces on an elastic halfspace (that is
translational invariance applies to two dimensions). Analyzing
\eq{eq:two_particle_interaction} then shows that $W$ has a pronounced
minimum for aligned force dipoles for all possible values of the
elastic constants \cite{uss:bisc03a,uss:bisc04a}.  Recently, such
alignment of cells in soft media has indeed been observed
experimentally \cite{c:vann03,c:engl04b}.

\begin{figure}
\begin{center}
\includegraphics[width=0.6\columnwidth]{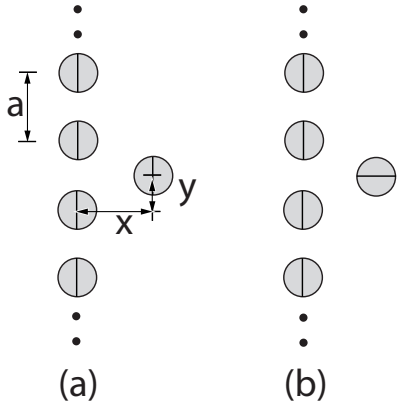}
\caption{Elastic interactions of cells lead to string
  formation. Here we consider the elastic interaction of a
  string of aligned dipoles spaced at equal distance $a$ with a single
  additional force dipole at horizontal distance $x$ and vertical
  offset $y$, both for (a) parallel and (b) perpendicular
  orientation.}
\label{Fig1}
\end{center}
\end{figure}
In the following we show that much physical insight into cellular
structure formation can be gained by starting from the finding that
elastic interactions of cells favor string formation.  We first
consider an infinitely extended string of aligned force dipoles spaced
at equal distance $a$. An additional dipole is placed at a horizontal
distance $x$ and with a vertical offset $y$, compare \fig{Fig1}.  To
simplify our notation, we use non-dimensional quantities: energy $W$
is rescaled with $P^2 / E a^3$ and length with $a$. Since the medium
is assumed to be linear, the superposition principle applies and the
effective interaction potential can be found by summing up all
pairwise interactions.  For parallel orientation (\fig{Fig1}a), we
have
\begin{equation} \label{eq:parallelstring}
W^{||} = - \partial^2_{y} \sum_{n=-\infty}^{+\infty}
\left(\frac{(1-\nu)^2}{\pi (x^2+(n-y)^2)^{\frac{1}{2}}}
+\frac{\nu (1+\nu) (n-y)^2}{\pi (x^2+(n-y)^2)^{\frac{3}{2}}}\right)
\end{equation}
and for perpendicular orientation (\fig{Fig1}b), we have
\begin{equation} \label{eq:orthostring}
W^{\bot} = - \partial_{x} \partial_{y} \sum_{n=-\infty}^{+\infty}
\frac{\nu (1+\nu) x (n-y)}{\pi (x^2+(n-y)^2)^{\frac{3}{2}}}.
\end{equation}
Eqs.~(\ref{eq:parallelstring}) and (\ref{eq:orthostring}) can be
analyzed further with methods from complex analysis. Briefly, the
Poisson sum formula $\sum_{n=-\infty}^{\infty}f(in) = 1/2i \int_C dz
\coth(\pi z) f(z)$ allows to turn the sums into integrals in the
complex plane \cite{e:alle04}.  By bending the contour $C$ around
suitable branch cuts introduced by the singularities of the
corresponding $f(z)$ at $z = \pm x + iy$, the integral forms of
Eqs.~(\ref{eq:parallelstring}) and (\ref{eq:orthostring}) can be
evaluated in the limit of large $x$:
\begin{equation} \label{eq:asymptotics}
W^{||/\bot} = 8 \pi \nu (1+\nu) \cos (2 \pi y) e^{- 2 \pi x} g^{||/\bot}(x)
\end{equation}
with
\begin{eqnarray} 
g^{||}(x) &=& -2 \pi \sqrt{x} + \frac{1}{\nu \sqrt{x}}
+ O\left(x^{-\frac{3}{2}} \right)\ , \\ 
g^{\bot}(x) &=& 2 \pi \sqrt{x} -\frac{1}{\sqrt x} 
+ O\left(x^{-\frac{3}{2}}\right)\ .
\end{eqnarray}
Thus, despite the long-ranged character of the elastic
pair-interaction, the effective interaction between an infinite string
and a single dipole (and a second string, respectively) is
short-ranged and decays to leading order $\sim
\sqrt{x} e^{-2\pi x}$. There is one exception, namely $\nu=0$,
when $W^{||} \sim e^{-2 \pi x}/\sqrt{x}$ for parallel dipoles and
$W^{\bot} = 0$ for perpendicular dipoles. In dimensional units, the
exponential decay occurs on a length scale $a/2\pi$ set by the dipolar
spacing $a$ only, independent of the elastic constants.

\begin{figure}
\begin{center}
\includegraphics[width=\columnwidth]{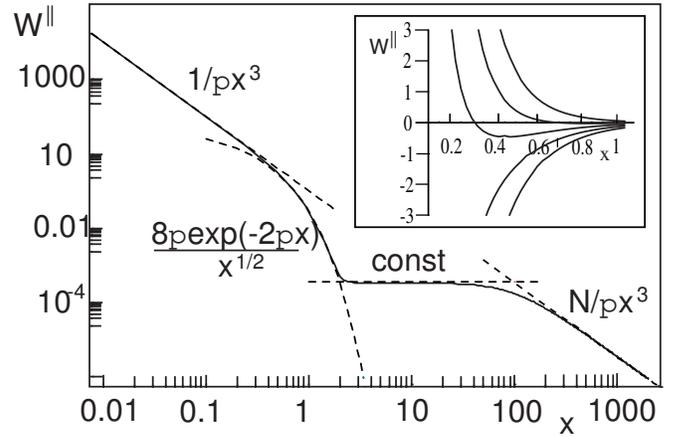}
\caption{Interaction $W^{||}$ between a finite string of $N$ aligned 
  dipoles and another dipole as a function of distance $x$ for
  $N=200$, offset $y=0$ and Poisson ratio $\nu=0$. The full line is
  the numerical evaluation of \eq{eq:parallelstring}. The dashed lines
  are the different scaling laws derived in the main text. Inset:
  $W^{||}$ for an infinite string for $x<1$, $y=0$ and
  $\nu=0,0.2,0.3,0.4,0.5$ from top to bottom.}
\label{Fig2}
\end{center}
\end{figure}
In practice, strings will be finite. Therefore we now consider a
single dipole with parallel orientation interacting with a finite
string of $N$ dipoles aligned along the y-axis and centered around the
origin.  For simplicity, we further specify to offset $y=0$ and
Poisson ratio $\nu=0$.  For $N=1$, the two dipoles interact via
$W^{||} = 1/(\pi x^3)$.  For $N=200$, the full interaction
potential as obtained by numerical evaluation of
\eq{eq:parallelstring} is shown as solid line in \fig{Fig2}.  Clearly
there are different scaling regimes, which can be explained in the
following way. For $x > N$, the dipoles in the string cancel each
other and the mechanical action of the string is equivalent to two
opposing forces $\pm P$ placed at $\pm N / 2$.  Therefore the string
effectively acts like one dipole of total strength $N P$ and the
asymptotic potential decays with the dipolar power law $N /(\pi x^3)$.
For $1 < x < N$, the string can be assumed to be infinite, but the
discrete spacing between the dipoles can still be neglected. Then we
can use the analogy to the electrostatic problem of an infinite,
homogeneously charged line. Using Gauss' law and the fact that the
force density vanishes gives that now the potential $W^{||}$ has to be
constant. The height of this plateau is fixed by the boundary
conditions, that is in our case by matching it to the large distance
regime at $x \sim N$; therefore $W^{||} \sim 1/N^2$.  For $x \approx
1$, the finite spacing between the dipoles becomes relevant and the
exponential decay $8 \pi e^{-2 \pi x} / \sqrt{x}$ predicted by
\eq{eq:asymptotics} becomes valid. Finally, for $x < 1$ the
interaction with the string is dominated by interactions with the
closest dipole in the string and $W^{||}$ crosses over to the dipolar
power law $1/(\pi x^3)$. \fig{Fig2} demonstrates the nice agreement
between this scaling analysis and the full numerical result.  While
these four scaling regimes are valid in general, the exact details are
very sensitive to dipole orientation, off-set $y$ and Poisson ratio
$\nu$. In particular, variations in $\nu$ can qualitatively alter the
interactions, especially for $x<1$, as demonstrated in the inset of
\fig{Fig2} for an infinite string.  In general, for an infinite string
the large distance regime and the height of the plateau vanish, and we
are left with the exponential decay derived above.

String formation is also known for passive particles, most prominently
for electric dipoles \cite{e:genn70,e:tlus00}.  The short-ranged
nature of the effective interaction between strings due to screening
inside the string is well-known in this case
\cite{e:made18,e:tao91,e:alle04}, but to our knowledge has not been
discussed before for force dipoles.  It has several important
implications for cellular structure formation.  First, it suggests
that long-ranged effects do not dominate structure formation at
particle densities sufficiently large as to allow formation of
strings of aligned dipoles.  Second, for high particle densities the
strong dependence of the string-string interaction on $\nu$ suggests
that structural changes might occur as a function of Poisson ratio.

\begin{figure}
\begin{center}
\includegraphics[width=\columnwidth]{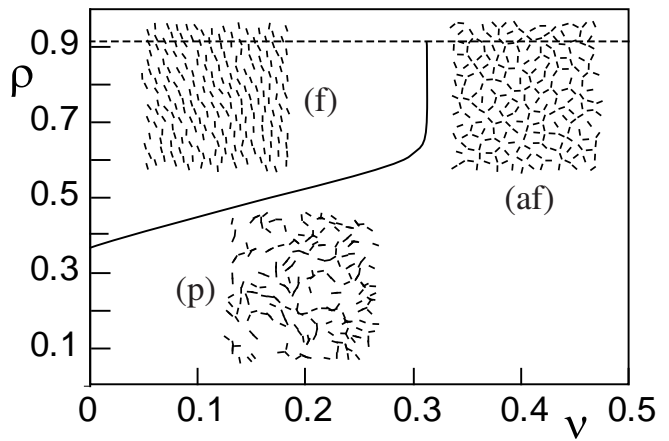}
\caption{Phase diagram for positionally disordered cells. At low cell density,
an orientationally disordered (\textit{paraelastic}) phase (p)
prevails. At high cell density, orientational order sets in, with a
nematic string-like (\textit{ferroelastic}) phase (f) at low values of
Poisson ratio $\nu$ and a isotropic ring-like
(\textit{anti-ferroelastic}) phase (af) at large values of $\nu$.}
\label{Fig3}
\end{center}
\end{figure}
In order to address these issues, we used Monte Carlo simulations to
study cellular structure formation on elastic substrates as a function
of reduced cell density $\rho$ and Poisson ratio $\nu$. Because cells
are supposed to interact only elastically, a circular exclusion zone
was attributed to each cell and positional degrees of freedom were
fixed at random, thus avoiding cell-cell contact (because we are in
two dimensions, $\rho \le 0.907$). We then relaxed the orientational
degrees of freedom using a Gibbs ensemble on $W$ with a small finite
effective temperature, which can be interpreted as an stochastic
element in cellular decision making.  Similar procedures have been
used before for modeling cellular structure formation
\cite{c:beys00}. Based on our simulations we predict three
different phases.  At low density, we always find many short strings
with hardly no correlation.  Presumably due to the screening effects
described above, no global order appears, except when an external
field is applied.  Therefore this phase might be termed
\textit{paraelastic}.  At high density and small values of the Poisson
ratio $\nu$, spontaneous polarization occurs. This phase of aligned
strings might be termed \textit{ferroelastic} and is characterized by
a non-vanishing nematic order parameter $S$.  As $\nu$ is increased at
high cell density, the nematic order parameter vanishes again for
$\nu_c > 0.32$. Now the local structure is ring-like rather than
string-like. This phase results from the Poisson effect and might be
termed \textit{anti-ferroelastic}. In \fig{Fig3}, we show typical
configurations and the phase transition line to the ferroelastic phase
as measured in simulations by setting $S=0.4$.

\begin{figure}
\begin{center}
\includegraphics[width=\columnwidth]{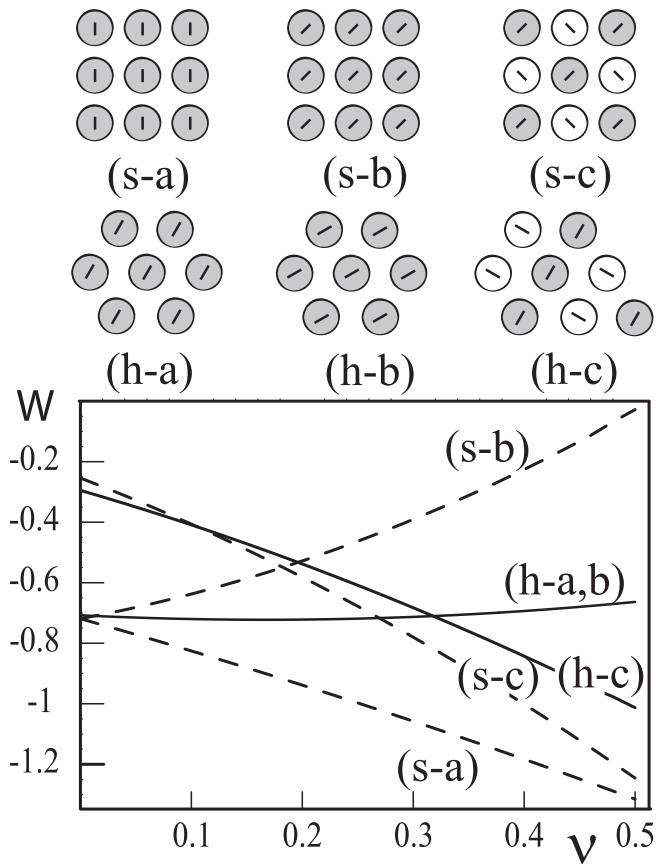}
\caption{Structure formation of cells positioned on square (s)
and hexagonal (h) lattices. Elastic interactions result in a
competition between ferroelastic (a,b) and anti-ferroelastic
structures (c) as a function of Poisson ratio $\nu$ and lattice
geometry.}
\label{Fig4}
\end{center}
\end{figure}
The important role of the Poisson ratio for cellular structure
formation can also be demonstrated for regular arrangements of force
dipoles. \fig{Fig4} shows six candidate structures identified by Monte
Carlo simulations for dipoles on square and hexagonal lattices.  To
identify optimal structures we calculate the interaction per
particle by evaluating the corresponding lattice sums. For this
purpose, we decompose the structures into A- and B-sublattices of
parallel strings, as indicated by grey and white colors.  The main
contribution to $W$ originates from interactions within a string and
is given by $-2 \zeta(3) (1+\nu)$, where $\zeta(z)$ is the Riemann
Zeta-function with $\zeta(3) \approx 1.29$.  Because the interaction
with adjacent strings decays exponentially as shown by
\eq{eq:asymptotics}, the interactions with adjacent strings are
dominated by interactions with the nearest neighbor strings and the
lattice sums converge quickly. The plots in \fig{Fig4} show that on
hexagonal lattices, a phase transition from a ferroelastic (h-a,b) to
an anti-ferroelastic structure (h-c) occurs at $\nu_c=0.32$. On square
lattices, $W$ decreases strongly with $\nu$ for the anti-ferroelastic
structure (s-c). However, here lattice geometry stabilizes the
ferroelastic phase (s-a) over the whole range of $\nu$, such that the
ferro-antiferroelastic transition does not occur. This finding also
implies that on substrates with $\nu \approx 0.4$ cellular structures
might be switched between ferro- and antiferroelastic phases simply by
varying the geometry of cell positioning.

In summary, we have shown that elastic interactions of cells lead to
surprising and non-trivial structure formation as a function of cell
density and Poisson ratio.  At low densities, the composite material
of cells and elastic medium (the tissue) behaves as a paraelastic
material: no global ordering occurs except for an externally applied
field.  Only for high cell density global order appears, but the
details depend strongly on Poisson ratio $\nu$. For small values of
$\nu$, a ferroelastic phase appears, that is a macroscopic anisotropic
force-dipolar field builds up that contracts the medium
unidirectionally. This situation is reminiscent of certain
pathological situations when wound contraction detoriates into
uncontrolled skin contraction (\textit{contracture}). For large values
of $\nu$, the Poisson effect leads to an anti-ferroelastic phase,
which is macroscopically isotropic, but which shows pronounced local
ordering into ring-like structures.

Elastic substrates appear to be the ideal experimental system to test
our predictions. However, the typical physiological environment of
tissue cells are the hydrated polymer networks of the extracellular
matrix and future work has to show how our results carry over to this
situation.  In particular, more details of the cellular decision
making process have to be included in this case.  Although then
isotropic linear elasticity theory is certainly not sufficient
anymore, one might expect that small values of the Poisson ratio $\nu$
correspond to cells probing single fibers (no Poisson effect), while
large values of $\nu$ correspond to cells probing more macroscopic
regions of extracellular space, which are incompressible due to
hydration effects.

We thank Phil Allen, Sam Safran and Assi Zemel for helpful
discussions. This work was supported by the Emmy Noether Program of
the DFG and the BIOMS Program at Heidelberg University.


\end{document}